# The Framework for the Prediction of the Critical Turning Period for Outbreak of COVID-19 Spread in China based on the iSEIR Model[1]

**George Xianzhi Yuan[1,2,3,4,5], Lan Di[6], Yudi Gu[7], Guoqi Qian[8] and Xiaosong Qian[2]**

[1]Business School, Chengdu University, Chengdu 610106 China

[2]Center for Financial Engineering, Soochow University, Suzhou 215008

Email: george_yuan99@suda.edu.cn (for Yuan) & qianxs@suda.edu.cn (for X Qian)

[3]School of Fintech, Shanghai Lixin University of Accounting & Finance, Shanghai 201209, China.

[4]Business School, Sun Yat-Sen University, Guangzhou 510275 China

[5]BBD Inc., Chengdu, China

[6]School of AI and Computer Science, Jiangnan University, Wuxi 214122, China. Email: dilan@jiangnan.edu.cn

7Center of Information Construct and Management, Jiangnan University, Wuxi, 214122， China. Email: udy1215@jiangnan.edu.cn

[8]School of Math & Statistics, University of Melbourne, Melbourne VIC 3010, Australia

Email: qguoqi@unimelb.edu.au

**Abstract**

The goal of this study is to establish a general framework for predicting the so-called critical "Turning Period" in an infectious disease epidemic such as the COVID-19 outbreak in China early this year. This framework enabled a timely prediction of the turning period when applied to Wuhan COVID-19 epidemic and informed the relevant authority for taking appropriate and timely actions to control the epidemic. It is expected to provide insightful information on turning period for the world's current battle against the COVID-19 pandemic.

The underlying mathematical model in our framework is the individual Susceptible-Exposed-Infective-Removed (iSEIR) model, which is a set of differential equations extending the classic SEIR model. We used the observed daily cases of COVID-19 in Wuhan from February 6 to 10, 2020 as the input to the iSEIR model and were able to generate the trajectory of COVID-19 cases dynamics for the following days at midnight of February 10 based on the updated model, from which we predicted that the turning period of CIVID-19 outbreak in Wuhan would arrive within one week after February 14. This prediction turned to be timely and accurate, providing adequate time for the government, hospitals, essential industry sectors and services to meet peak demands and to prepare aftermath planning.

Our study also supports the observed effectiveness on flatting the epidemic curve by decisively imposing the "Lockdown and Isolation Control Program" in Wuhan since January 23, 2020. The Wuhan experience provides an exemplary lesson for the whole world to learn in combating COVID-19.

---

[1] **The corresponding author is: George Yuan with email: george_yuan99@suda.edu.cn.**

## 1. The Background and Related Literature

Infectious disease epidemics always present challenges to human society, threatening the safety of human life and causing social upheaval and economic losses. In recent years, novel virus outbreaks have been increasing worldwide, from the 2003 SARS-CoV, the H1N1 influenza A virus in 2009, the MERS-CoV in 2012, the Ebola virus in 2015, the Chai virus in 2016, the H5N7 avian influenza virus in 2017, to the recent coronavirus (COVID-19) that emerged at the end of 2019 (in Wuhan, China). These outbreaks have brought great loss to human life, disrupted population processes, and negatively impacted global development.

The current coronavirus disease 2019 (COVID-19) outbreak began in Wuhan City (Hubei province of China) in later December 2019 and quickly spread to other cities in China in a matter of days. It was announced as a public health emergency of international concern by the World Health Organization (WHO) on January 30, 2020. Predicting the development of the development of the outbreak as early and as reliably as possible is critical for action to prevent its spread with necessary implementation of emergency response plan.

Furthermore, the current COVID-19 outbreak is a particularly urgent public health event. Apart from the challenges the virus presented to China's health system, it has spread rapidly to other regions and countries worldwide. We currently do not have a unanimous metric of estimating when the global situation of the virus will be under control.

In this study, we like to discuss how to establish a general framework for the prediction of the critical so-called "Turning Period" which would play a very important role in assisting better plans for the time frame of emergence plans, in particular for associated looking forward planning such as the battle with the current pandemics of COVID-19 worldwide.

By assessing the performance of our iSEIR model to predict the timeline of the spread of COVID-19 in Wuhan since late December 2019 under the help of our new concept of "Turning Time Period (Time Period)" to predict the control of the epidemic outbreak measured by a reduction in the number of people infected, it shows that our iSEIR model (see Yuan [28] for more in details and the brief introduction given below), an extension of the SEIR model, describes the spread and behavior of infectious diseases for individuals under the framework of a probability perspective works very well to accurately predicted that "the COVID-19 situation in China would peak around mid- to late February as early as February 7, 2020".

Indeed, we like to emphasize that the identification of the Turning Time Period is the key to have a successful implementation for emergency plan as it provides a timeline for effective actions and solutions to combat a pandemic by reducing as much unexpected risk as soon as possible.

Our study also indicates that the implementation of the emergency program in the practice associated with the "Isolation Control Program (or, say, Wuhan Quarantine Program (see Begley [36])) " since January23, 2020 by China in national level may be a good experiences by other countries and regions to take a lesson.

During almost last century, in the study and modelling mechanics of infectious diseases, the traditional model called "SEIR" denoted for " infectious disease dynamics susceptible–exposed–infectious–resistant" and its various (deterministic) versions have been introduced and been very popular in analyzing and predicting the development of an epidemic (see Liu et al. [1], Murray [2], Wu et al. [3-4], Prem et al. [5], Li et al.[6], Lin et al.[7], Kuniya [8], Roosa et al. [9] and references wherein). The SEIR models the flows of people between four states: susceptible state variable "S", exposed variable "E", infected variable "I", and resistant variable "R". Each of those variables represents the number of people in those groups. Take COVID-19 as one example, assume that the average number of exposed cases that are generated by one infected person of COVID-19, this number could be regarded as the so-called "**basic reproduction number**" (which is indeed the expected number of cases directly generated by one case in a population where all individuals are susceptible to infection), the study on the basic reproduction number, related features for the globally stable endemic and disease-free equilibria and thresholds is always the main stream for people from the academic research community to the practice in the subject of epidemic disease spread behavior and related social issues.

In particular, a great deal and effort have been done for the study on the process and evolution of the limits of the Basic Reproduction Number and similar thresholds in predicting global dynamics of epidemics. In particular, since the occurrence of COVID-19 later December, 2019 in Wuhan, the study on the impact such as how serious the outbreak of infectious disease to the society, and to prediction how many people would be infected to become infectious, and so on, have been attracted by a large number of scholars with reports, see Cao et al. [10-11], Cowling and Leung [12], Hermanowicz [13], Li et al. [14], Guan et al. [15] and reference wherein.

On the other hand, modelling the situation of COVID-19 and effects of different containment strategies in China with dynamic differential equations and parameters estimation have also been paid a lot attention by a number of scholars, e.g., see Gu et al. [17], Hu et al. [18], Zhao et al. [18], Yan et al. [20], Wang et al. [21], Tang et al. [22], Huang et al [23], Cui and Hu [24] and related references wherein.

In particular, Professor Murray (see [2]) leads his IHME COVID-19 health service utilization forecasting team to work on the estimates of predicted health service utilization and deaths due to COVID-19 by day for the next 4 months for each state in the US. Their objective is to determine the extent and timing of deaths and excess demand for hospital services due to COVID-19 in the US (also, see the study of Kuniya [8] on Japan, Murray [2] on USA, Wu et al. [3-4] , Prem et al. [5] on China).

It seems that almost all of them still follow the way to pay the attention mainly on modelling or forecasting the behavior of spread for epidemic disease directly related to those infected who also become infectious, i.e., the variable "I" of SEIR model. No study pays the attention on the study how to establish a general framework for the prediction of the critical turning period for the spread of pandemic diseases (e.g., the outbreak of COVID-19) in general.

The object of this paper is to fill in this gap as we do believe that it is so important to study the general dynamics for the outbreak of COVID-19 in each country or regions, which faces one simple expectation that how to find in which time period the battle with COVID-19 will be under controlled ? By a simple fact that for any spread of infectious disease, we know that in general it is impossible to find or identify the exact turning point (or critical point) for a big pandemic (which means the behavior of the spread of COVID-19 virus is under control) due to various dynamics and associated uncertainty! But if using the idea to think of the a time period (or say, time interval) instead of an exact time point to identify the true change for the behavior of spread for epidemic disease such as in terms of the number of infectious people have significantly be reduced, plus the population of the exposed (E) are also true under the control to reach to certain low level, then it seems possible for us to identify different phases and stages for the mechanics of the outbreak of infectious disease incorporating with some useful tools such as the iSEIR one we introduced in [28].

Thus the goal of this paper is to discuss how to build the framework for the prediction of the Critical Turning Period for Outbreak of pandemics (e.g., the spread of COVID-19) based on the application of our iSEIR Model. It expects that the concept for the prediction of the critical turning period would provide us a better way to prepare the emergency plan for the prevention and control of COVID-19 in the practice, such as working on the estimates of predicted health service utilization and deaths due to COVID-19 by day for the next 4 months for each state in the US (see [2] for more information), countries and regions worldwide.

By assessing the performance of prediction by using the iSEIR model for the timeline of the spread's mechanics of COVID-19 in Wuhan on dates of Feb.6 and Feb.10, 2020 by using the concept of "Turning Time Period (Time Period)" to forecast the time frame for the control of the epidemic outbreak measured by a reduction in the number of people infected, it shows that our iSEIR model (an extension of the SEIR model) works very well to accurately predicted that "the COVID-19 situation in China would peak around mid- to late February as early as February 7, 2020". This review also shows that the identification of the Turning Time Period is the key to have a successful implementation for emergency plan as it provides a timeline for effective actions and solutions to combat a pandemic by reducing as much unexpected risk as soon as possible.

Our study also indicates that the implementation of the emergency program in the practice associated with the "Isolation Control Program (or, say, Wuhan Quarantine Program ) since January23, 2020 by China in national level may be a good experiences by other countries and regions to take a lesson.

This paper consists of 6 sections as follows.

## 2. The Challenges faced by the Emergency Mechanism of Epidemic Prevention and Control of Infectious Diseases Worldwide Today

The idea of the key "**SEIR** Epidemic Model" can be traced back to Dr. Ronald Ross who received the Nobel Prize for Physiology or Medicine in 1902 for his work on malaria which laid the foundation for combatting the epidemic disease (see [28]). In 1927, Kermack and McKendrick formulated a simple deterministic model called SIR to describe the dynamic mechanism for directly transmitted viral or bacterial agent in a closed population (see [26]). Since then, scholars have contributed and advanced this field; a significant milestone in the study of Epidemics was the publication of "The Mathematical Theory of Infectious Diseases" in 1957 by Bailey (see [27]). Of these, the famous SEIR model, a core subject in Epidemic discipline (see [3]), like mentioned above, is the basis for describing the mechanism for the spread of infectious diseases, and has been used in a number of research projects and related applications. In the SEIR model, the "S" state refers to the susceptible group (or ignorants) who are susceptible to disease but have not been infected yet; the "E" state refers to the exposed group who are infected but are not infectious yet (or lurkers); the "I" state refers to those infected who also become infectious; and the "R" state refers to those who have recovered from the infection (through treatment or natural recovery) who may or may no longer be infectious, or those who have passed away.

### 2.1 The Challenges faced by the Emergency Mechanism of Epidemic Prevention and Control

Infectious diseases have always been a major challenge to human society, threatening the safety of human life and causing social upheaval and economic losses. Every scenario of an epidemic outbreak due to a novel infectious disease carries a similar set of challenges: the unknown nature of the new pathogen/strain, a lack of immediate effective treatment and vaccine, and an ill-prepared public health infrastructure to accommodate the surge in potential patients and need for testing. Public health policies that could alleviate and help prevent the impact and scale of an outbreak require significant and massive governmental and societal implementation of emergency planning and intervention strategies. At present, I want to focus the objective of this paper towards **three key issues** in approaching these outbreaks:

1) **How do we establish a spatiotemporal model for the infectious diseases' outbreak:** in order describe the mechanics of the spread of infectious diseases?
2) **How do we conduct numerical simulation and risk prediction indicators**: in order to conduct numerical simulation based on the real scenes, which can be used to provide an outlook and planning schedule associated with a key period known as the "**Turning Phase**" during the spread of infectious diseases?
3) **How do we carry out effective predictive analysis on the epidemic situation of infectious diseases on an ongoing basis:** in order to cooperate with dynamic management, support public health emergency plans/services, and support community responses by establishing a coherent bigdata method for data fusion from different sources with different structure.

In combatting these outbreaks, the immediate implementation of an emergency response mechanism delays an epidemic's peak which affords us more time to control the epidemic by reducing the number of infections in a concentrated period of time. Thus, a successful emergency plan lengthens the "Turning Phase" (or say, "Turning time Period" (see [29], [30] and also [34]), the time interval between $T_0$ and $T_1$. Effective ways of flattening the curve include intervention actions such distancing or isolation programs (e.g., quarantine program (see [33]-[35] and [36]).

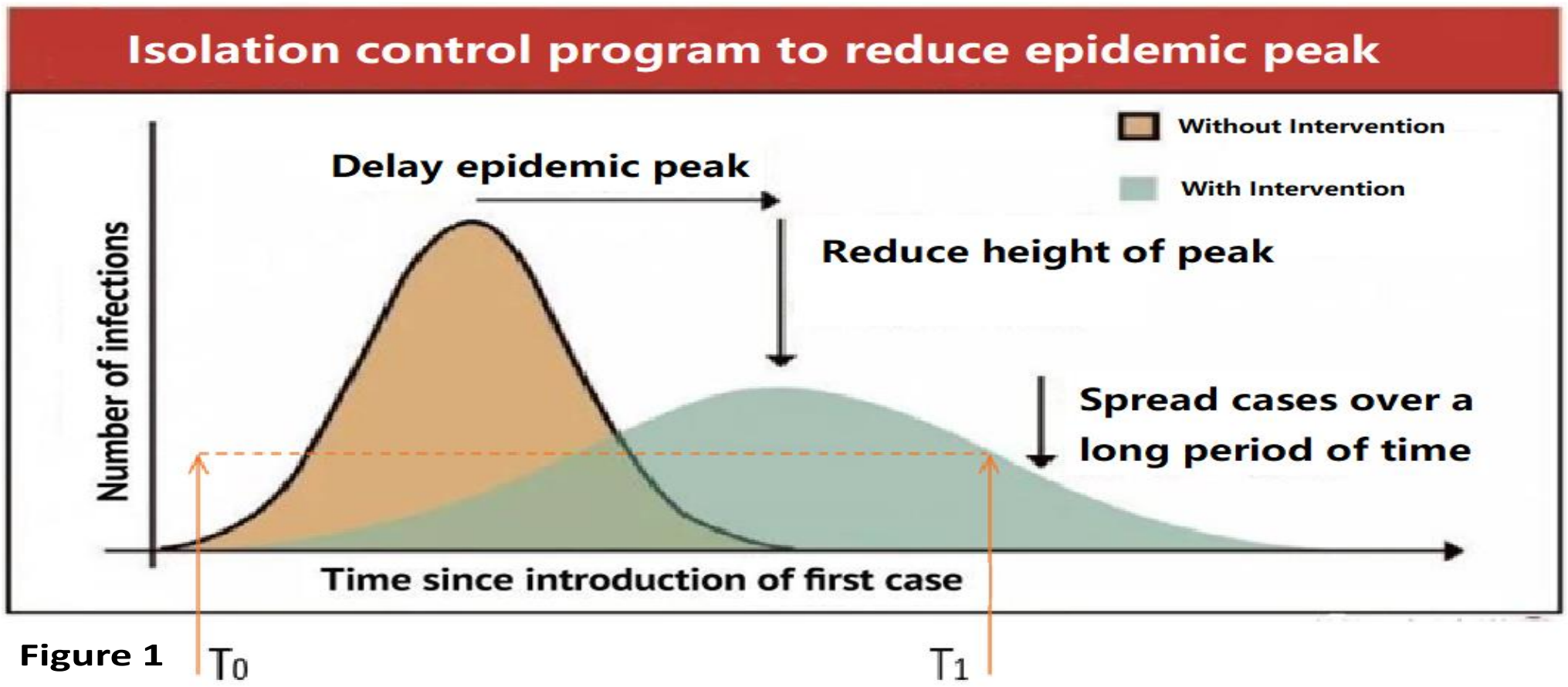

Figure 1

Thus, a major challenge faced by our current responses to epidemic prevention and infectious disease control is finding a way to predict the critical time period "**Turning Period**" or "**Turning Phase**" when implementing an emergency plan. Knowing this timeline is critical to combat the outbreak of an epidemic or infectious disease.

In this section, we discuss the framework for predicting the "**Turning Phase**" using our **iSEIR Model** introduced in [28] which was used to successfully predict the "Turning Phase" for the outbreak of epidemic COVID-19 virus from January 2020 to early of March 2020 in China using only data from Feb. 10, 2020. Here we give a brief introduction for our **iSEIR mode**, which stands for the name "**individual Susceptible-Exposed-Infective-Removed**".

## 2.2 The Framework of our iSEIR Dynamic Model with Multiplex Networks

For the convenience of our discussion, we give an introduction on the general framework of our iSEIR model which was introduced in [28] for more in details and numerical simulations lined to the applcations.

In one word, our iSEIR model operates under a probability perspective for each individual with the name "**individual Susceptible-Exposed-Infective-Removed (iSEIR)**", which is an extension of the classic SEIR one. The iSEIR model allows us to conduct simulations from the individual levels located on the nodes of different community networks by incorporating its uncertainty with the probability to conduct random simulation in the corresponding multiplex network.

### 2.2.1 The Classical SEIR Model

For the SEIR model, the state S refers to the susceptible group (or ignorants) who are susceptible to the disease but have not been infected yet; state E refers to the exposed group who are infected but are not infectious yet; state I refers to those infected who also become infectious; and state R refers to those who have recovered from the infection (through treatment or natural recovery) and are no longer infectious. We also use S(t), E(t), I(t) and R(t) to represent the proportion of the population being in state S, E, I and R at time t, respectively. In the present case, the system of ODEs describing the dynamics of an SEIR epidemic model thus has the following form:

$$\begin{cases} \frac{dS}{dt} = -\mu\langle k\rangle SE \\ \frac{dE}{dt} = \mu\langle k\rangle SE - \beta EI \\ \frac{dI}{dt} = \beta EI - \lambda I \\ \frac{dR}{dt} = \lambda I \end{cases} \quad (1)$$

where μ is the rate at which an exposed individual becomes infective, λ is the recovery rate, and with normalization condition (each variable is in percentage change)

$$S(t)+E(t)+I(t)+R(t)=1 \quad (2)$$

for every $t \geq 0$ (since the population is considered closed).

The above deterministic SEIR model and its generalizations have received a lot of attention from various researchers. Indeed, the SEIR model represents more accurately the spread of an epidemic than the corresponding SIR model that does not take into account the latent period. The SEIR model has a slower growth rate, since after the pathogen invasion the susceptible individuals need to pass through the exposed class before they can contribute to the transmission process, as shown (by Figure I_SEIR model illustration) below:

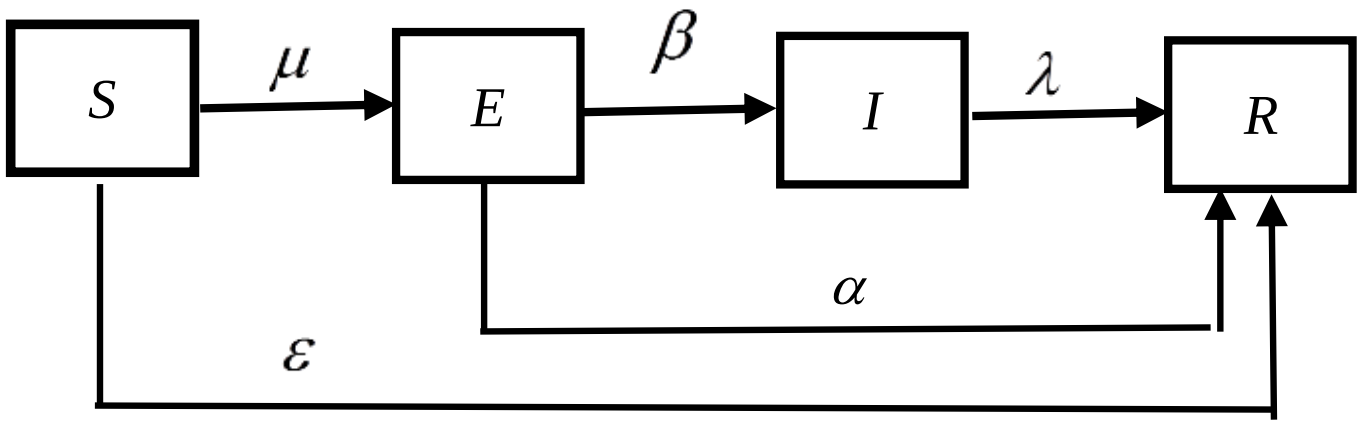

Figure I_SEIR model-Illustration

### 2.2.2 The Framework of Our iSEIR Dynamic Systems with Multiplex Networks

Based on the former discussion, the iSEIR (see Figure_II_iSEIR model Illustration below) is an extension of SEIR model, but presented in different expression, which is a component form as follows in Eq.(3).

$$
\begin{cases}
\dfrac{dS}{dt} = A - S(t)E(t)\sum_i \mu_i \sum_j p_{ij} - S(t)\sum_i \varepsilon_i \\
\dfrac{dE}{dt} = S(t)E(t)\sum_i \mu_i \sum_j p_{ij} - E(t)I(t)\sum_j \beta_j \sum_k q_{jk} - E(t)\sum_j \alpha_j \\
\dfrac{dI}{dt} = E(t)I(t)\sum_j \beta_j \sum_k q_{jk} - I(t)\sum_k \lambda_k \\
\dfrac{dR}{dt} = S(t)\sum_i \varepsilon_i + E(t)\sum_j \alpha_j + I(t)\sum_k \lambda_k
\end{cases}
\tag{3}
$$

where, parameters in the systems are illustrated as following:

- where A is the growth rate of new arrivals.
- $\mu_i$ denotes the transfer probability from S(t) to E(t).
- $p_{ij}$ denotes the connection probability of the i-th sample in S(t) to the j-th sample in E(t); it equal to 1 if connected or 0 if not.
- $\varepsilon_i$ denotes the transfer probability from S(t) to R(t), which is the removed probability.
- $\beta_j$ denotes the transfer probability from E(t) to I(t).
- $q_{jk}$ denotes the connection probability of the j-th sample in E(t) to the k-th sample in I(t); it equal to 1 if connected or 0 if not.
- $\alpha_j$ denotes the transfer probability from E(t) to R(t).
- $\lambda_k$ denotes the transfer probability from I(t) to R(t).

The proposed iSEIR model is shown as in Figure II_iSEIR model Illustration below:

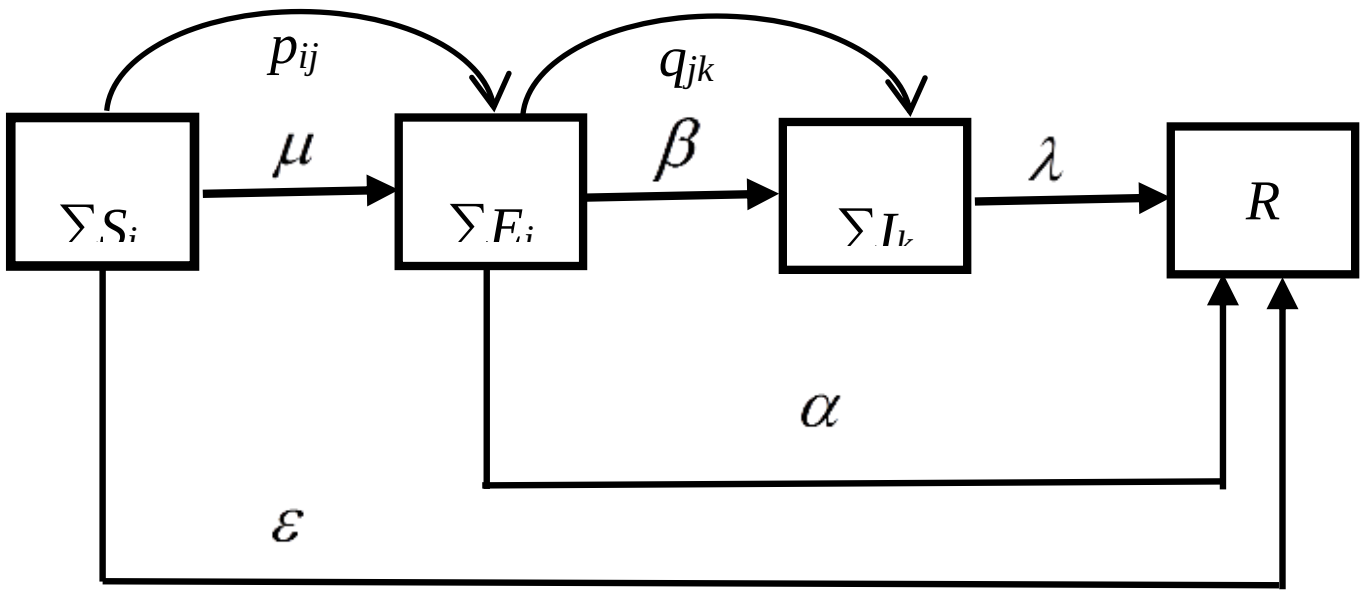

Figure II_iSEIR model Illustration

In order to make the system (3) short and analyze easily, we use the following denotation:

$$
\begin{cases}
\mu = \sum_i \mu_i; \ \varepsilon = \sum_i \varepsilon_i; \ \alpha = \sum_j \alpha_j; \ \beta = \sum_j \beta_j; \ \lambda = \sum_k \lambda_k \\
p = \sum_i \sum_j p_{ij}; q = \sum_j \sum_k q_{jk}
\end{cases}
\tag{4}
$$

As our iSEIR model is based on the framework of multiplex networks, we also give some notations here, more in details are given in [28] (also see notations' specification listed for the simulations based on iSEIR given by **Appendix I** below).

By suppose the population S for the spreading consists of N individuals $S_j$, j =1, ..., N; namely S: ={ $S_j$, j = 1, ..., N }, and also suppose these N individuals are distributed over M continuous domains $U_i$, i = 1,……, M , where a domain may refer to a residential district or a network. Then we can conduct the simulation based on framework of probability respective for each individual based on iSEIR in a population multiplex network by following five steps:

**Step 1:** We first allow the transition from state S to state R directly with probability ε per unit time (the same below) by following equation:

$$\begin{cases} \frac{dS}{dt} = A - \mu SE - \varepsilon S \\ \frac{dE}{dt} = \mu SE - \beta EI - \alpha E \\ \frac{dI}{dt} = \beta EI - \lambda I \\ \frac{dR}{dt} = \varepsilon S + \alpha E + \lambda I \end{cases} \quad (5)$$

where A is the growth rate or new comer; $\varepsilon$ is the probability of a susceptible person being directly transformed into an immune person by means of e.g., isolation; μ is the rate of a susceptible being infected; β is the rate of an infected person becoming infectious; α is the rate of an infected person becoming immune directly; and λ is the rate of an infectious person entering into an immune state. Figure 1 below gives a visual presentation of (5). Note that there is no direction transition between S and I in (5) as illustrated by Figure 1 above.

**Step 2:** Each individual in the rumor spreading network at time t is identified by its state and position in that state group.

**Step 3:** We will establish an adjacency matrix to describe the influence effects between individuals.

**Step 4:** Computing the probabilities of transitions between states involves considering the following two aspects (the K-adjacency method).

**Step 4.1:** the distances between uninfected individuals and their neighborhoods of infected individuals within; and

**Step 4.2:** the number of individuals infected.

**Step 5:** The full specification of the model is given by combining steps 1 to 4 together with an individual-level representation of (5) that is illustrated in Figure 2 above.

These five steps will help us to run the simulations for the observation of the so-called "Supersaturation Phenomenon" under the probability perspective of all individuals by applying iSEIR mode which help us to identify the "Turning Phase", which are critical for any emergency plan being successful by responding to the challenge any outbreak of pandemics under the emergency in the practice. Thus the iSEIR will be used as a tool for us to discuss how we can establish the framework for the prediction of the critical "Turning Phase" for the emergency implementation response in an epidemic infectious disease outbreak described in next section.

## 3. The Framework for the Prediction of the Critical "Turning Phase" based on our iSEIR Model for the Emergency Implementation Response in an Epidemic Infectious Disease Outbreak

In this section, we discuss the framework for predicting the "**Turning Phase**" using our **iSEIR Model** introduced in [26] which was used to successfully predict the "Turning Phase" for the outbreak of epidemic COVID-19 virus from January 2020 to early of March 2020 in China using only data from Feb. 10, 2020. As previously discussed, in the battle for epidemic prevention and control of infectious disease outbreak, it is crucial to implement effective prevention measures in the early stages of an outbreak. Furthermore, identifying the beginning and ending points of the time interval which forms the "**Turning Time Period** (**Time Period**)" lets us know how long to expect to implement emergency protocols to effectively flatten the curve. It is possible to make this predictive analysis of the "Turning Time Period" using the iSEIR model through the "**Supersaturation Phenomenon**" elaborated below (see [28]).

### 3.1 The Concept of "Turning Time Period (Turing Period)" for Turning Phase

While the exact turning point (or critical point) for any infectious disease spread cannot be precisely determined due to various dynamics and associated uncertainty, by borrowing the variables of "**Delta**" and "**Gamma**" practiced in financial risk management (see Hull [37]) it is possible to identify the upper and lower limits using, for example, the current tolerable degrees from the change in new (confirmed) patients daily. These indicators then allow us to identify the different phases and stages of an infectious disease outbreak through the iSEIR model, which we elaborate in Part A and Part B. Through this method, it is possible to calculate the critical time period of the "Turning Time Phase (Turning Phase)."

**Part A: To Identify Different Time Phases of Epidemic Infectious Diseases Spread**

We propose the identification of three general three phases (time periods) for the emergency response of Epidemic Infectious Diseases Spread paired with medical response actions as elaborated below:

1) **The First Phase**: The initial starting stage corresponds to the initial occurrence and prepare for possible emergence plan of a new virus which may or may not transform into a new epidemic.
2) **The Second Phase**: For our consideration, this is the most important phase which is the so-called "**First Half-Time Phase**" otherwise known as the "**Turning Phase**" (or "**Turning Period**") which starts with the beginning of a possible outbreak and includes the delayed epidemic peak by implementation of emergency planning to control the disease spread. The First Half Time phase involves Delta and Gamma indicators (elaborated more in Part B below) to measure the daily change in number of new patients (i.e., the indicator "Delta"), and the rate of the daily change in number of new patients (i.e., the indicator "Gamma"). As shown in Figure 1 above, the time interval from $T_0$ to $T_1$ is our Turning Phase (Turning Period), and also makes the end of the "First Half Time Phase" for an epidemic infectious diseases spread.
3) **The Third Phase**: During this stage, the epidemic infectious disease spread enters the do-called "**Second Half -Time Phase**", which means the epidemic peak is gone and the rate of spread is under greater control. This is measured in a continuously decreasing rate of new infections per day, and ultimately leads to any but not exclusively of the following scenarios: a): the disease completely disappears; b): an effective vaccine/treatment is introduced; and c): the strain could also disappear and reappear cyclically in seasons, or other reasons.

Of the three phases, the most important time period to identify is the beginning and the ending time points of the **First-Half Phase** (known as the "**Turning Period**"). This phase is crucial to controlling the outbreak and spread of an epidemic infectious disease after the first case of occurrence.

Thus, being able to identify the "**First Half-Phase**" is crucial for the reliable prediction of the "**Turning Period**" (or "**Turning Phase**") as the ending time point of the Turing Period will allow us to predict when the outbreak of the infectious disease is under the control by the level we may settle (incorporating with ability and capability in the practice).

The next challenge to address is how to identify or predict this Turning Period.

To determine the Turning Period, we look to the occurrence of the so-called "Supersaturation Phenomenon" (elaborated below) based on our iSEIR model (see [28]-[31] and also the report by [34]) by running the simulation for the four control variables S(t), E(t), I(t) and R(t) in the iSEIR model (see [28]). These variables are functions of time "t" under the probability framework of individuals involved in the epidemic disease's spread. Our prediction can be achieved once we observe the so-called "**supersaturation phenomenon**", **the moment where the future value range of $T_1$ observes both E(t) and I(t) decrease**. We determine this by simulating an iSEIR model that incorporates data from the initial daily disease spread (further explained by Part B below) (see also Figure 3 for both E(t) and I(t)) going down).

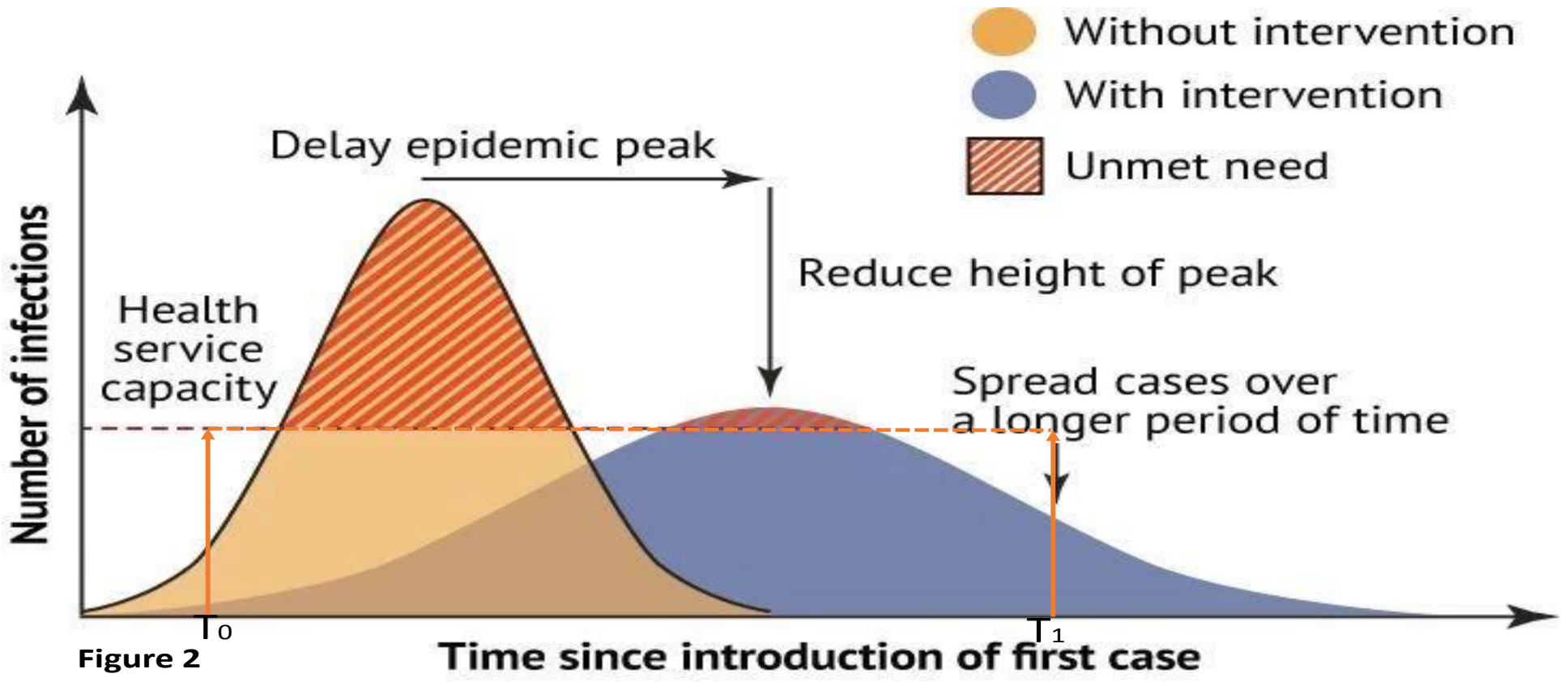

**Figure 2**

**Part B: The Prediction of the Turning Period by using iSEIR Model Associated with Delta and Gamma Risk Indicators**

When COVID-19 first emerged, we suggested using Delta and Gamma indicators to measure the daily change of new patients and the change rate of the daily change for new patients. These variables allowed us to identify the beginning point of the time interval for the Turning Period within China. For example, to identify the starting point $T_0$ of the Turning Period, the level we considered for Delta was settled as **no greater than 20% daily**; and Gamma as **no larger than 1% daily**.

Next, to predict the future ending time point $T_1$ of the Turning Period, which measures when the epidemic disease spread is under our control, we run numerical simulations of our iSEIR model as shown by Figure 3. $T_1$ is determined to be the point at which both control variables E(t) and I(t) drop.

The combination of Part A and Part B allowed us to reliably predict the Turning Period of COVID-19 since its first case in Wuhan (China) in late December 2019. Using only the available data released by the National Health Commission of China from Feb. 10, 2020, we correctly assessed that "**COVID-19 peak around mid- to later February, and entering in the Second Half Period around Feb. 20, 2020**." ( see our report [31] and confirmed by WHO's report [34]).

To further understand our predictive simulation, we briefly describe the new idea of our iSEIR model and the related "Supersaturation Phenomenon".

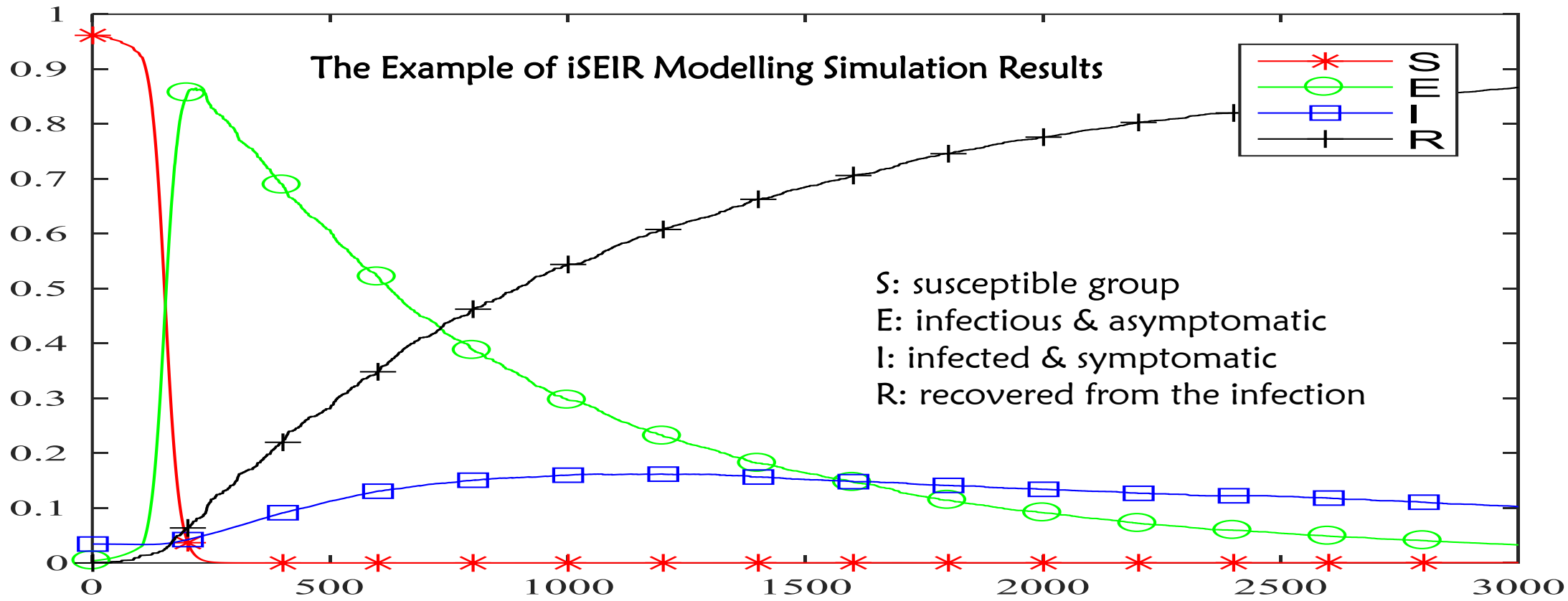

**Figure 3** iSEIR Modelling Simulation

where the vertical y-axis represents a standardized unit (normalized unit, it is explained as the range from 0 to 1), the horizontal x-axis represents time, a unit is "10 minutes".

### 3.2 Our iSEIR Model as a new tool to model for "Supersaturation Phenomenon"

Almost all key models such as SEIR and related mathematical tools that exist to model the mechanics of epidemic disease spread are established under deterministic frameworks which assume that all individual behaviors and patterns are uniform (i.e., all behaviors of individuals are homogeneous), but this is not true as each individual's behavior of infecting or being infected is different. In order to have a better way of describing the dynamics of "**Spreading behavior**" in multiplex network at an individual level to community to population levels, we introduced the so-called **iSEIR model (individual Susceptible-Exposed-Infective-Removed)** which operates under a probability perspective for each individual (see reference [28]) around two years ago as an extension of the classic SEIR one. This **iSEIR** model allows us to conduct simulations from the individual level located on the nodes of different community networks by incorporating its uncertainty with the probability to conduct random scenarios study with consideration of the corresponding multiplex networks. The behavior distribution of S, E, I and R can also be numerically simulated from the **iSEIR** model with properly specified values of parameters on population scales (in the change of percentage), population density, and transfer rate and so to have the simulation results given by Figure 3 above.

Thus, the simulation results based on the iSEIR model suggest that the intensity and extensiveness of the spread of the disease can be lowered by external intervention under a so-called "**supersaturation phenomenon.**" This phenomenon occurs when at some point in the future (denoted by $T_1$) when both of the variables "E(t)" and "I(t)" drop in value and do not increase anymore as shown around the x-axis value of 1500 units in Figure 3.

At this value, both variables "E(t)" and "I(t)" are at the "supersaturation phenomenon." Thus, the "supersaturation phenomenon" in the iSEIR model allows us to predict the turning period for the outbreak of epidemic diseases spread in the application.

## 4. The Study of our Prediction for the "Peak Period" of COVID-19" in China for Early February 2020

The methodology outlined above is what we expect when modeling epidemic disease, including COVID-19 which has become a global pandemic since the case of its first outbreak in Wuhan in late December 2019.

Taking into account the performance of intervention controls (i.e., Wuhan quarantine program in Hubei province (see [35])) implemented since Jan. 23, 2020 at a national level by the Chinese government, we predicted the "Peak Time Period" of COVID-19 in China using our iSEIR model (see [29]-[32]) in two separate reports on Feb. 6 and Feb. 10, 2020 (see [33]-[35]).

First of note, as we discussed above, we truly believe that "**Turning Period**" (or say, "**Turning Phase**") for **COVID-19** is not a single time point, but a period of time interval associated with at least the two following measures, "Delta" and "Gamma" explained below:

**4.1 The Indicators of Delta and Gamma used to identify the Beginning Turning Point for the " Turning Period" of COVID-19 in China was February 1, 2020.**

We borrow the concepts of "Delta" and "Gamma" which are called Greek Letters used in financial risk management: "Delta" measures the daily change of both "E" and "I"; and "Gamma" accounts for the speed of daily change for the number of new confirmed patients for both "E" and "I". In order to predict the Turning Phase for COVID-19 in China, we need to first identify the "Starting Point" of COVID-19's Turning Period.

By using the daily information from Jan. 23, 2020, to Feb. 6, 2020, we first observed that "Delta" and "Gamma" were below 20% and 2% daily, respectively, **for 5 consecutive days since February 1,** which led us to conclude that Feb. 1 heralds the start towards the situation peaking, and thus we first have the following conclusion (see [29] with more in details).

**Conclusion A:**

Feb. 1, 2020, is the starting time point for all of China (including Wuhan, Hubei) and the beginning Turning Point for the " Turning Period" (see report [5] released on Feb. 7, 2020) of COVID-19 as verified by the official daily data released by NHC of China on Feb. 28, 2020 (see also reports from [31]-[32] and [33]-[34]).

Indeed, as the data shown below, we observed that "Delta" and "Gamma" were below 20% and 2% respectively for 5 consecutive days since February 1. This led them to conclude that February 1 was the starting point of the situation peaking.

**Table 1: The Daily Change of Infected COVID-19 Patients by The National Health Protection Committee of China from Feb. 1 to Feb. 6 2020/表 1: 中国国家卫健委新冠状肺炎疫情动态数据汇报**

| Times/时间 | Feb 1/20 Turning point / 拐点开始 | Feb. 2/2020 | Feb/ 03/2020 | Feb.04/ 2020 | Feb. 05/2020 | Feb.06/2020 |
|---|---|---|---|---|---|---|
| COVID-19 Infected Patients in Hubei / 湖北确诊感染人数 | 9,074 | 11,177 | 13,522 | 16,678 | 19,665 | 22,112 |
| COVID-19 Infected Patients in rest of China / 其他省份确诊感染人 | 5,306 | 6,028 | 6,916 | 7,646 | 8,353 | 9,049 |
| China COVID-19 Infected Patients in China / 全国确诊感染人数 | 14,380 | 17,205 | 20,438 | 24,324 | 28,018 | 31,161 |
| Delta: Daily Change of COVID-19 Infected Patients in Hubei / 湖北确诊感染人（日）百分比变化 | 27% | 23% | 21% | 23% | 18% | 12% |
| Delta: Daily Change of COVID-19 Infected Patients in rest of China / 其他省份感染人数（日）百分比变化 | 14% | 14% | 15% | 11% | 9% | 8% |
| Delta: Daily Change of COVID-19 Infected Patients in China / 全国新增确诊感染人数的变化 | 22% | 20% | 19% | 19% | 15% | 11% |
| Gamma: Speed of Daily Change of COVID-19 Infected Patients in Hubei / 湖北新增确诊感染人数（日）的增减速度 | 16% | -14% | -9% | 11% | -23% | -31% |
| Gamma: Speed of Daily Change of COVID-19 Infected Patients in rest of China / 其他省份新增确诊感染人数变化的增减速度 | -25.6% | -2.8% | 8.3% | -28.3% | -12.4% | -9.9% |
| Gamma: Speed of Daily Change of COVID-19 Infected Patients in China / 全国新增确诊感染人数日度变化的速度 | 1% | -11% | -4% | 1% | -20% | -26% |

### 4.2 The iSEIR model as a tool to predict that "the COVID-19 situation in China would peak around mid- to late February 2020" on February 10, 2020.

Using the concept of Turning Phase and the application of the iSEIR model on February 10, 2020, based on 3 weeks of available daily data (from Jan. 23, 2020 to Feb. 10,2020) released by NHC of China, our simulation successfully modeled the future pattern of official data released almost 3 weeks later, as shown by the conclusion below (the reports from [29]-[30], the corresponding simulations' inputs by applying iSERI model given by **Appendix I** with incorporation of the official data released by NHC of China as of Feb. 10, 2020).

**Conclusion B:**

**First**, with February 1, 2020, as the initial starting point of the turning point of the **COVID-19** epidemic situation in China, combined with the simulation analysis of our internal model iSEIR in considering the current situation and the "**isolation control program"** currently implemented in China, the time interval for controlling the COVID-19 outbreak should be "**in mid-February, before the end of February 2020"** (see also the report released by [29]-[30] for more details).

The above conclusion was reached by the following simulation analysis based on the iSEIR model on Feb. 10, 2020 with the reports given by [29] & [30]:

The iSEIR model calculated that at the national level (excluding Hong Kong, Macao and Taiwan) as of Feb. 10, 2020 that (see also results from simulations given by Figure below):

1. The number of infectious and asymptomatic (E) (in % change) will reach a pe ak in about 2 days (Feb. 12, 2020), and then after about 7 days (Feb. 17, 202 0), the rate of increase will decrease to less than 10%;
2. The daily increase in the number of people infected and symptomatic (I) is a steady downward trend, and the rate will decrease to within 10% after about 14 days (Feb. 24, 2020);
3. The proportion of people recovering from infection (R) will return to more tha n 80% after about 17 days (Feb. 27, 2020).

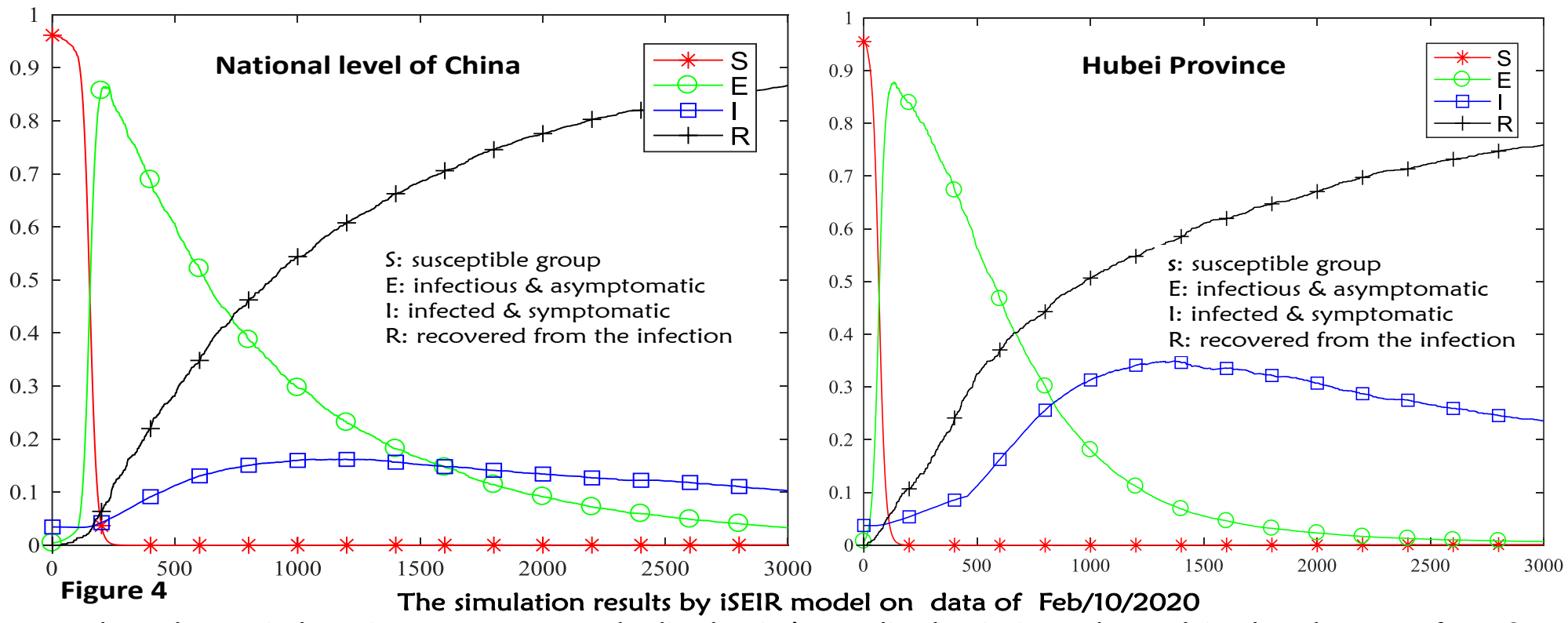

**Figure 4** The simulation results by iSEIR model on data of Feb/10/2020
where the vertical y-axis represents a standardized unit (normalized unit, it can be explained as the range from 0 to 1), the horizontal x-axis represents time, one unit is "10 minutes".

**Second**, as a result, we earmarked the second half of February (around Feb. 20, 2020) as the period when the peak would happen as shown by our simulation results shown below on Feb. 10, 2020 (later confirmed by the data released by NHC of China).

Our conclusion that "the number of infectious and asymptomatic (E) (in % change) will reach a peak in about 2 days (Feb. 12, 2020)" was confirmed by the following data released by the National Heath Commission (NHC) of China :

The daily official data shows that Feb. 11, 2020 was the peak time with its Delta value (the daily change for the number of close contacts people (with patients)) at 5.4% daily, the highest value in the date range from Feb. 1 through to Feb. 25. This corroborates our prediction of around Feb. 12, 2020 as the peak based on our iSEIR model.

Table 2: The Daily Data of COVID-19 China

| Item/项目 | 2020/2/19 | 2020/2/18 | 2020/2/17 | 2020/2/16 | 2020/2/15 | 2020/2/14 | 2020/2/13 | 2020/2/12 | 2020/2/11 | 2020/2/10 | 2020/2/9 | 2020/2/8 | 2020/2/7 |
|---|---|---|---|---|---|---|---|---|---|---|---|---|---|
| The number of Close contracts/ 正在接受医学观察 | 126,363 | 135,881 | 141,552 | 150,539 | 158,764 | 169,039 | 177,984 | 181,386 | 185,037 | 187,728 | 187,518 | 188,183 | 189,660 |
| Its Delta | -7.0% | -4.0% | -6.0% | -5.2% | -6.1% | -5.0% | -1.9% | -2.0% | -1.4% | 0.1% | -0.4% | -0.8% | 1.9% |
| Irs gamma | 74.8% | -32.9% | 15.2% | -14.8% | 20.9% | 168.0% | -4.9% | 37.6% | -1380.0% | -131.7% | -54.6% | -140.1% | -1271.8% |
| | | | | | | | | | | | | | |
| The number of Close contracts/ 追踪到密切接触者 | 589,163 | 574,418 | 560,901 | 546,016 | 529,418 | 513,183 | 493,067 | 471,531 | 451,462 | 428,438 | 399,487 | 371,905 | 345,498 |
| Its Delta | 2.6% | 2.4% | 2.7% | 3.1% | 3.2% | 4.1% | 4.6% | 4.4% | 5.4% | 7.2% | 7.4% | 7.6% | 10.0% |
| Its Gamma | 6.5% | -11.6% | -13.0% | -0.9% | -22.5% | -10.7% | 2.7% | -17.3% | -25.8% | -2.3% | -3.0% | -23.7% | -9.2% |

### 4.3 The Prediction is "Turning Period, not Turning Point" in Supporting the Emergency Response Plan in Application

Based on study above, we would like to note that our iSEIR model does not predict a specific date for the peak of pandemic virus but rather an approximate interval encompassing a larger time period. We believe that the so-called 'Turning Point' is not an accurate term; it is not a single time point, but rather a time interval period.

This "Turning Period" of the latter half of February is backed by the World Health Organization (WHO) whose own data shows that the Chinese cases of COVID-19 levelled off sometime during the week of February 14. WHO director-general Tedros Adhanom Ghebreyesus also stated in a press conference on February 24 that the epidemic in China "has been declining steadily" since February 2, 2020 (see [33]-[35]).

In a follow-up analysis to determine the accuracy of our ISEIR model using official data from February 11–21, 2020, we produce the following findings (see also report by [32-33]):

1. Since February 17, the "Gamma" of the number of close contacts has declined rapidly to 4% (Table3 below);
2. Since February 17, the death rate has remained within 3% (Table4 below);
3. Since February 17, the "Gamma" of the number of infected people has also de clined to within 2–4% (Table5 below).

In addition to the fact that February 17–18 had the highest number of confirmed cases in China (Table6), "we have reason to believe that the inflection point has emerged on February 17 and 18, thus validating the prediction we made on February 7" (see the "statement 2." of the findings above) (the report given by [29-32]):

| Table3: People with close contact with patients /表 3：追踪密切接触者（累计值） | National /全国 | Daily Change /日度变化 | Hubei Prov. /湖北省 | Daily Change /湖北日度变化 |
|---|---|---|---|---|
| 2020/2/12 | 471,531 | 4.3% | 158,377 | 3.9% |
| 2020/2/13 | 493,067 | 4.4% | 166,818 | 5.1% |
| 2020/2/14 | 513,183 | 3.9% | 176,148 | 5.3% |
| 2020/2/15 | 529,418 | 3.1% | 183,183 | 3.8% |
| 2020/2/16 | 546,016 | 3.0% | 191,434 | 4.3% |
| 2020/2/17 | 560,901 | 2.7% | 199,322 | 4.0% |
| 2020/2/18 | 574,418 | 2.4% | 206,087 | 3.4% |
| 2020/2/19 | 589,163 | 2.6% | 214,093 | 3.9% |
| 2020/2/20 | 606,037 | 2.9% | 225,696 | 5.4% |
| 2020/2/21 | 618,915 | 2.1% | 234,217 | 3.8% |
| 2020/2/22 | 628,517 | 1.6% | 240,937 | 2.9% |
| 2020/2/23 | 35,531 | 1.1% | 246,781 | 2.4% |
| 2020/2/24 | 41,742 | 1.0% | 251,265 | 1.8% |
| 2020/2/25 | 47,406 | 0.9% | 255,750 | 1.8% |
| 2020/2/26 | 52,174 | 0.7% | 259,491 | 1.5% |
| 2020/2/27 | 56,054 | 0.6% | 262,195 | 1.0% |
| 2020/2/28 | 58,587 | 0.4% | 263,916 | 0.7% |
| 2020/2/29 | 60,716 | 0.3% | 265,617 | 0.6% |

| Table4:Case fatality rate (cumulative) /表 4: 病死率（累计值） | National /全国病死数 | Daily change /全国百分百 | Hubei Province /湖北病死数 | Daily Change in Hubei/湖北百分百 |
|---|---|---|---|---|
| 2020/2/12 | 1,259 | 2.29% | 1,202 | 2.55% |
| 2020/2/13 | 1,380 | 2.16% | 1,318 | 2.54% |
| 2020/2/14 | 1,523 | 2.29% | 1,457 | 2.68% |
| 2020/2/15 | 1,665 | 2.43% | 1,596 | 2.84% |
| 2020/2/16 | 1,770 | 2.51% | 1,696 | 2.91% |
| 2020/2/17 | 1,868 | 2.58% | 1,789 | 2.98% |
| 2020/2/18 | 2,004 | 2.70% | 1,921 | 3.11% |
| 2020/2/19 | 2,118 | 2.84% | 2,029 | 3.25% |
| 2020/2/20 | 2,236 | 2.96% | 2,144 | 3.40% |
| 2020/2/21 | 2,345 | 3.07% | 2,250 | 3.55% |
| 2020/2/22 | 2,442 | 3.17% | 2,346 | 3.66% |
| 2020/2/23 | 2,592 | 3.36% | 2,495 | 3.88% |
| 2020/2/24 | 2,663 | 3.43% | 2,563 | 3.96% |
| 2020/2/25 | 2,715 | 3.48% | 2,615 | 4.01% |
| 2020/2/26 | 2,744 | 3.50% | 2,641 | 4.03% |
| 2020/2/27 | 2,788 | 3.54% | 2,682 | 4.07% |
| 2020/2/28 | 2,835 | 3.58% | 2,727 | 4.11% |
| 2020/2/29 | 2,870 | 3.60% | 2,761 | 4.13% |

| Table5: Confirmed Patients (cumulative value) /表 5：确诊病例（累计值） | National /全国 | Daily Change /全国日度变化 | Hubei Province /湖北 | Daily Change of Hubei Province/湖北日度变化 |
|---|---|---|---|---|
| 2020/2/12 | 58,761 | 31.6% | 47,163 | 41.4% |
| 2020/2/13 | 63,851 | 8.7% | 51,986 | 10.2% |
| 2020/2/14 | 66,492 | 4.1% | 54,406 | 4.7% |
| 2020/2/15 | 68,500 | 3.0% | 56,249 | 3.4% |
| 2020/2/16 | 70,548 | 3.0% | 58,182 | 3.4% |
| 2020/2/17 | 72,436 | 2.7% | 59,989 | 3.1% |
| 2020/2/18 | 74,182 | 2.4% | 61,682 | 2.8% |
| 2020/2/19 | 74,576 | 0.5% | 62,031 | 0.6% |
| 2020/2/20 | 75,891 | 1.8% | 63,088 | 1.7% |
| 2020/2/21 | 76,288 | 0.5% | 63,454 | 0.6% |
| 2020/2/22 | 76,936 | 0.8% | 64,084 | 1.0% |
| 2020/2/23 | 77,150 | 0.3% | 64,287 | 0.3% |
| 2020/2/24 | 77,658 | 0.7% | 64,786 | 0.8% |
| 2020/2/25 | 78,064 | 0.5% | 65,187 | 0.6% |
| 2020/2/26 | 78,497 | 0.6% | 65,596 | 0.6% |
| 2020/2/27 | 78,824 | 0.4% | 65,914 | 0.5% |
| 2020/2/28 | 79,251 | 0.5% | 66,337 | 0.6% |
| 2020/2/29 | 79,824 | 0.7% | 66,907 | 0.9% |

**Table6:The change in existing confirmed infected patients in China /表 6：全国现存确诊病例变化时间表**

| Time/时间 | The existing confirmed infected patients of China /全国确诊病例存量 | The change of National Patients /全国确诊病例存量 | The existing confirmed infected patients of Hubei /湖北确诊病例存量 | The change of Hubei Patients /湖北日度变化 |
|---|---|---|---|---|
| 2020/2/11 | 38,800 | | 29,659 | |
| 2020/2/12 | 51,860 | 33.7% | 42,789 | 44.3% |
| 2020/2/13 | 55,748 | 7.5% | 46,806 | 9.4% |
| 2020/2/14 | 56,872 | 2.0% | 48,175 | 2.9% |
| 2020/2/15 | 57,416 | 1.0% | 49,030 | 1.8% |
| 2020/2/16 | 57,927 | 0.9% | 49,847 | 1.7% |
| **2020/2/17** | **58,016** | **0.2%** | 50,338 | 1.0% |
| **2020/2/18** | 57,802 | -0.4% | **50,633** | **0.6%** |
| 2020/2/19 | 56,727 | -1.9% | 50,091 | -1.1% |
| 2020/2/20 | 55,389 | -2.4% | 49,156 | -1.9% |
| 2020/2/21 | 53,285 | -3.8% | 47,647 | -3.1% |
| 2020/2/22 | 51,606 | -6.8% | 46,439 | -5.5% |
| 2020/2/23 | 49,824 | -6.5% | 45,054 | -5.4% |
| 2020/2/24 | 47,672 | -7.6% | 43,369 | -6.6% |
| 2020/2/25 | 45,604 | -8.5% | 41,660 | -7.5% |
| 2020/2/26 | 43,258 | -9.3% | 39,755 | -8.3% |
| 2020/2/27 | 39,919 | -12.5% | 36,829 | -11.6% |
| 2020/2/28 | 37,414 | -13.5% | 34,715 | -12.7% |
| 2020/2/29 | 35,329 | -11.5% | 32,959 | -10.5% |

## 4.4 Reviewing our Analysis based on the iSEIR Model

Here we review the results based on our iSEIR model from Feb. 10, 2020. We present the original report in two parts as follows.

### 4.4.1 The Outlook Report for COVID-19 as of February 10, 2020 (see [30] & [31])

Based on our iSEIR model and the data from China's novel COVID-19 epidemic released as of Feb. 10, 2020 ([30] & [31]], we reach the following conclusions I, II, and III (and the corresponding inputs for the simulations by applying iSEIR mode are given by **Appendix I** below and incorporation with official data released by NHC of China on Feb.10, 2020).

**I): COVID-19 on the National Level of China**

Using 428,438 close contacts people, 42,638 confirmed infected people and 2.38% confirmed death rate as the key input parameters on Feb. 10, 2020, our iSEIR simulation analysis results show (see the figure below, where the vertical y-axis represents one unit of Standardization (from 0 to 1) and the horizontal x-axis represents Time whereby one unit is "10 minutes," that is, one day is "144 units"):

1) On the national level, the change in number of close contacts (E) reached a peak in about 2 days (around Feb.12, 2020), and then after about 7 days (i.e., around Feb.17, 2020), the number of close contacts began a stable downward trend of less than 10%;
2) On the national level, the daily decrease in the number of people infected (I) is also a steady downward trend as of date, and is within 10% after approximately 14 days (Feb.24, 2020);
3) On the national level, the proportion of people recovering from infection (R) will quickly return to around 80% after about 17 days (around Feb. 27, 2020).

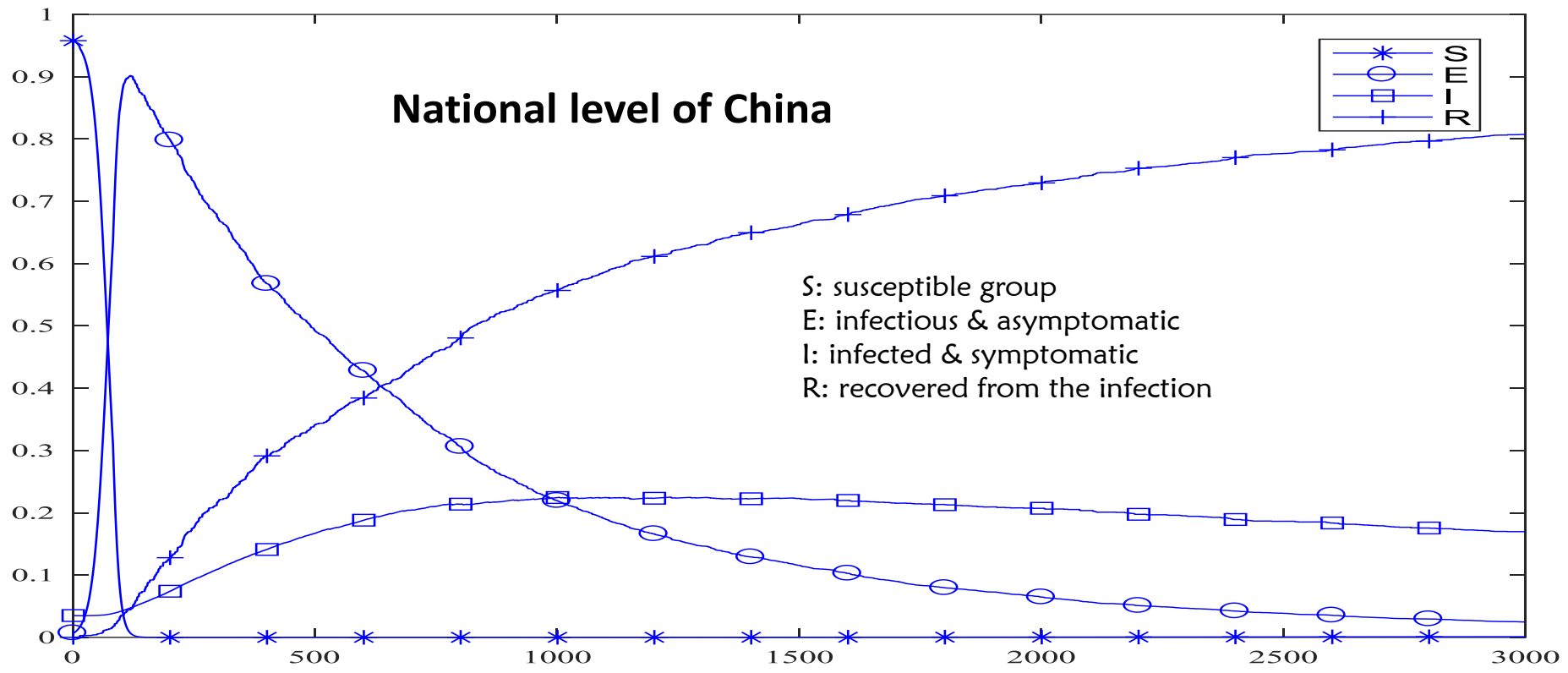

**Figure 5** The simulation results by iSEIR model on data of Feb/10/2020
where the vertical y-axis represents a standardized unit (normalized unit, it can be explained as the range from 0 to 1), the horizontal x-axis represents time, one unit is "10 minutes".

**II): COVID-19 on the Level of Hubei Province**

By using 144,279 close contacts, 31,728 confirmed infections and 3.07% mortality as the core input parameters, as of Feb. 10, 2020, our simulation analysis results show (see the figure below):

1) In Hubei Province, the number of close contacts (E) peaks after about 2 days (around Feb. 12, 2020), and then steadily declines with a change of less than 5% after about 6 days (around Feb. 16, 2020);
2) In Hubei Province, the daily decrease of the number of people infected (I) also shows a steady downward trend, reaching a peak in about 7 days (Feb. 17, 2020), and then decreasing at a rate of about 7%;

3) In Hubei Province, the proportion of people who recovered from infection (R) quickly recovers to around 70% after about 17 days (around Feb. 27, 2020).

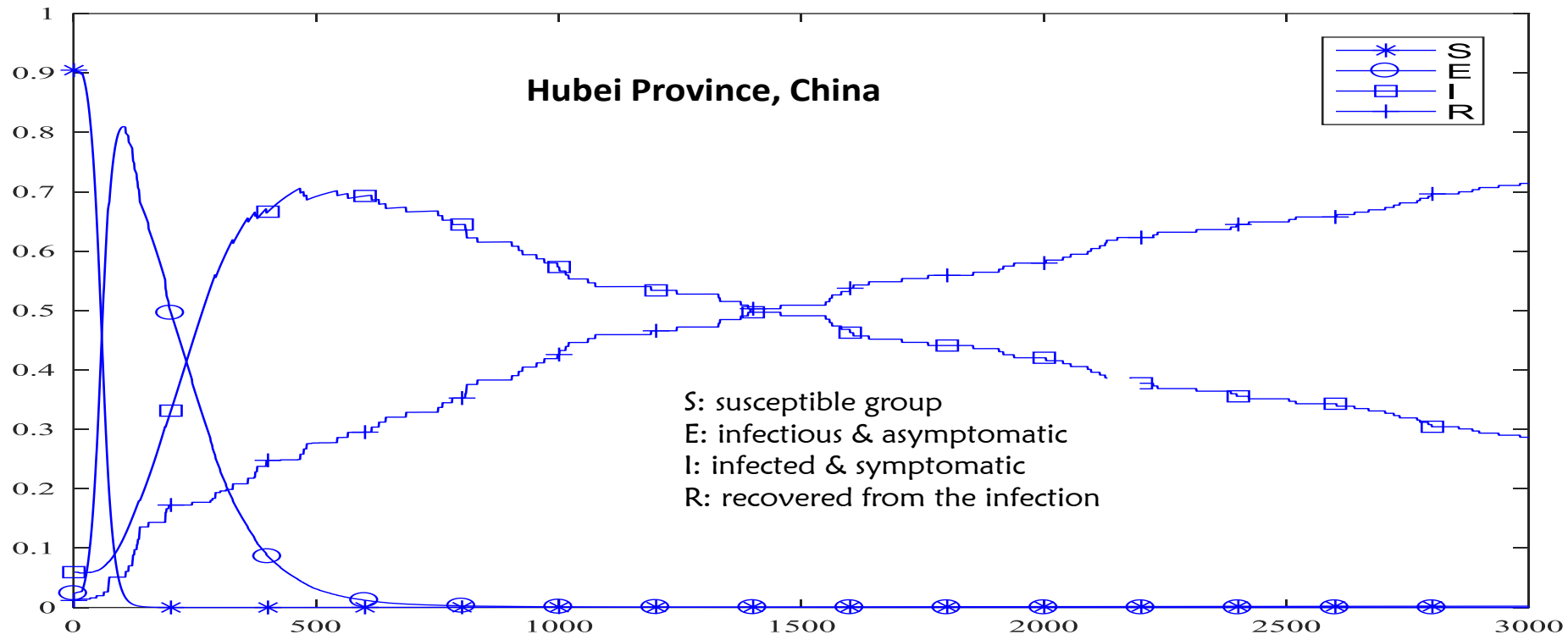

**Figure 6** The simulation results by iSEIR model on data of Feb/10/2020

where the vertical y-axis represents a standardized unit (normalized unit, it can be explained as the range from 0 to 1), the horizontal x-axis represents time, one unit is "10 minutes".

## III): COVID-19 on the Level of Wuhan City

Using 83,917 close contacts (estimated), 18,454 confirmed infections, and 4.05% mortality rate as the key input parameters as of Feb. 10, 2020, combined with the current intervention situation in Wuhan, our simulation analysis results show (see the figure below):

1) In the city of Wuhan, after about 4 days, the daily variation of the number of close contacts (E) is a stable downward range of about 3%;
2) In the city of Wuhan, the daily decrease of the number of people infected (I) is also a steady downward trend, reaching a peak in about 4 days, and then decreasing at a rate of about 4%;
3) In the city of Wuhan, the proportion of people (R) recovering from infection quickly recovers to close to 70% after about 17 days (around Feb. 27, 2020).

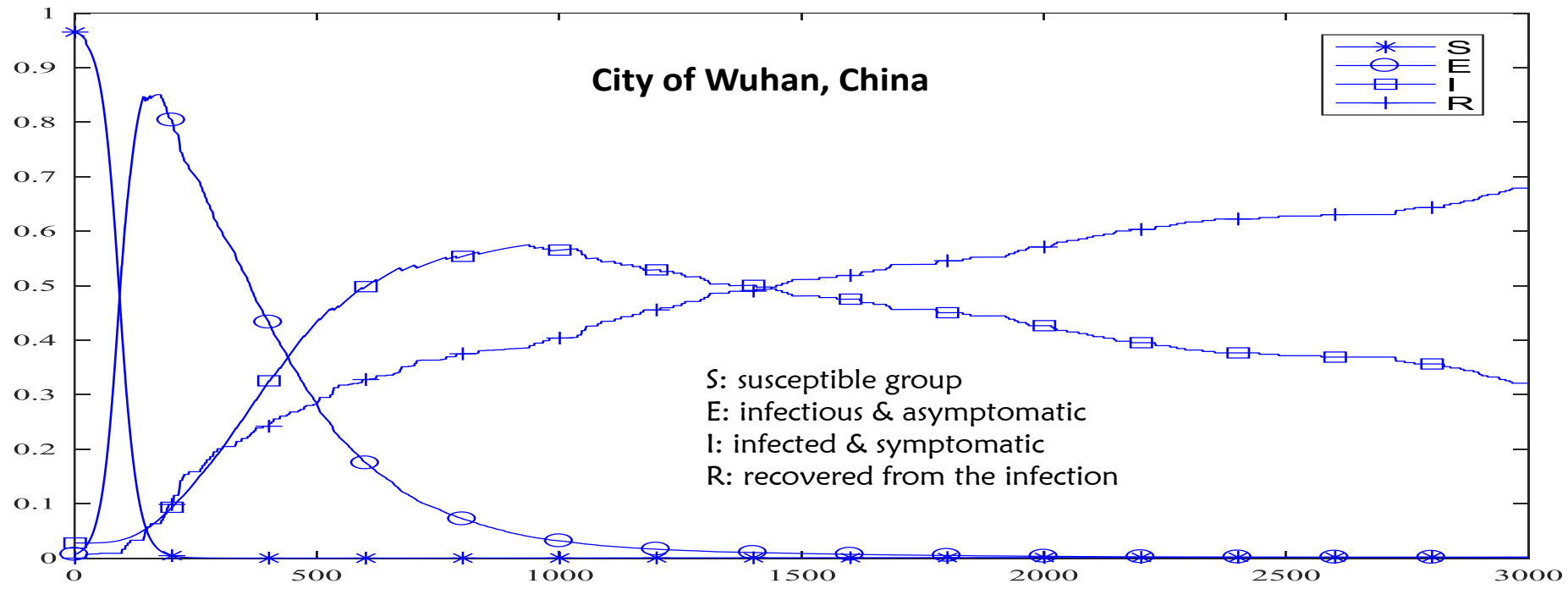

**Figure 7** The simulation results by iSEIR model on data of Feb/10/2020
where the vertical y-axis represents a standardized unit (normalized unit, it can be explained as the range from 0 to 1), the horizontal x-axis represents time, one unit is "10 minutes".

### 4.4.2 Entering into the Second Half of the COVID-19 Outbreak on February 20, 2020

On Feb. 16, 2020, we released the report (see [31]-[32]) that, "Once the peak has been identified, a different approach has to be taken with regard to city management, economic development and epidemic control measures." This is in-line with President Xi Jinping's remarks on Feb. 23, where he stressed the need to resume production and normal life.

Since the COVID-19 virus could be spread asymptomatically and does not yet have an official vaccine, being able to model its trend is crucial to timeline projections. Dr. Bruce Aylward, an epidemiologist who led a 25-member team from WHO to conduct a 9-day field study trip to Beijing, Guangdong, Sichuan and Hubei, shared this viewpoint. "They are using big data, artificial intelligence in places," Dr. Aylward said, "It's a technology-powered and science-driven agile response at a phenomenal scale."(see [32]-[34]) and also the report [35]).

## 5. What We learned from COVID-19 in China

Our predictions based on the iSEIR model on Feb. 6, 2020, were validated by the official data released by the NHC on COVID-19. The numbers of China and its six regions (including Hubei Province and Hubei's five cities: Wuhan, Xiaogun, Suizhou, Huanggang, and Huangshi) were as follows: 76,936 confirmed patients in national level of China as of Feb. 22 (in which 64,084 confirmed cases from Hubei province) and 12,852 in other provinces, 62,8517 people with close contacts (with patients), and a 3.17% mortality rate of confirmed infected. Our conclusions, originally published on Feb. 16, 2020, are as follows (see [31-34]) stated by the following two parts.

**The First**:

1) February 1, 2020, is the starting point for the formation of the epidemic's Turning (Inflection) Point which will form around mid-February (see [29] for the concept and definition of "Turning (Inflection) Point" formation and the establishment of supporting standard indicators);
2) We reported that the COVID-19 viral outbreak in China has reached controllable measures in our February 16, 2020, report, and that we are now entering the second-half of the struggle against COVID-19 in China (see [30]-[32]).

**The Second:**

There appears the following similar development trends in Hubei Province, as well as in Wuhan, Xiaogan, Suizhou, Huanggang and Huangshi (with reference to the figure below):

1) The number of close contacts (represented by "E") begins to decline rapidly after about February 16. Then, in a short period of time, the number of close contacts follows a stable downward change of less than 10%;
2) The daily variation of change of new confirmed infected (represented by "I") is also in a stable downward variation. In a very short time (no more than 3 to 10 days), the number of daily change for infected people is within the range of 10% (but higher than the overall average).

Using the official data of the epidemic situation released as of Feb. 21, 2020, by the National Health Commission (NHC) of China, the following three tables show the daily changes in the number of close contacts, the number of infections, and the death rate of confirmed patients in five cities (Wuhan, Xiaogan, Suizhou, Huanggang and Huangshi). The official data confirms the predictive analysis reports we made on Feb. 6, and Feb. 10, 2020. Based on this, we find that:

1) Since Feb. 17, the daily variation of close contacts in all seven simulated areas has declined rapidly to 4% (the national average change is lower than that in Hubei Province);
2) Since Feb. 17, the death rate of confirmed patients in China has remained within 3%, but the death rate of confirmed patients in Hubei Province is slightly higher than that in China, which is remains at about 3.5%;

3) Since Feb. 17, the daily variation of the number of infected people in all seven simulated areas has also declined to within 4% (the daily variation of Wuhan is slightly higher than that of the whole country, but that of Xiaogan, Suizhou, Huanggang and Huangshi remains within 2%).

| Tabke7: Track close contacts with patients (cumulative)/表8：追踪密切接触者（累计值） | National #/全国 | Daily Change National/全国日度变化 | Hubei #/湖北 | Daily Change in Hubei/湖北日度变化 |
|---|---|---|---|---|
| 2020/2/12 | 471,531 | 4.3% | 158,377 | 3.9% |
| 2020/2/13 | 493,067 | 4.4% | 166,818 | 5.1% |
| 2020/2/14 | 513,183 | 3.9% | 176,148 | 5.3% |
| 2020/2/15 | 529,418 | 3.1% | 183,183 | 3.8% |
| 2020/2/16 | 546,016 | 3.0% | 191,434 | 4.3% |
| **2020/2/17** | **560,901** | **2.7%** | **199,322** | **4.0%** |
| **2020/2/18** | **574,418** | **2.4%** | **206,087** | **3.4%** |
| **2020/2/19** | 589,163 | **2.6%** | 214,093 | **3.9%** |
| **2020/2/20** | 606,037 | **2.9%** | 225,696 | **5.4%** |
| **2020/2/21** | **618,915** | **2.1%** | **234,217** | **3.8%** |
| **2020/2/22** | **628,517** | **1.6%** | **240,937** | **2.9%** |

| Table8: The fatality rate (cumulative)/表8：病死率（累计值） | National # of death/全国病死数 | Ratio in National/全国百分百 | Death # in Hubei/湖北病死数 | Ratio in Hubei湖北百分百 |
|---|---|---|---|---|
| 2020/2/12 | 1,259 | 2.29% | 1,202 | 2.55% |
| 2020/2/13 | 1,380 | 2.16% | 1,318 | 2.54% |
| 2020/2/14 | 1,523 | 2.29% | 1,457 | 2.68% |
| 2020/2/15 | 1,665 | 2.43% | 1,596 | 2.84% |
| 2020/2/16 | 1,770 | 2.51% | 1,696 | 2.91% |
| **2020/2/17** | **1,868** | **2.58%** | **1,789** | **2.98%** |
| **2020/2/18** | **2,004** | **2.70%** | **1,921** | **3.11%** |
| 2020/2/19 | 2,118 | **2.84%** | 2,029 | **3.25%** |
| 2020/2/20 | 2,236 | **2.96%** | 2,144 | **3.40%** |
| **2020/2/21** | **2,345** | **3.07%** | **2,250** | **3.55%** |
| **2020/2/22** | **2,442** | **3.17%** | **2,346** | **3.66%** |

**Therefore, we arrive at the following general conclusions:**

1) Based on the number of existing confirmed cases, we determined the inflection point to be around Feb. 17, 2020. Based on the number of existing confirmed cases in China and Hubei Province, we used the iSEIR model on Feb. 11, 2020, to predict that Feb. 17, and Feb. 18, 2020 would have the highest number of existing confirmed cases in China and Hubei Province respectively. Today (as of Feb. 22, 2020), the number of existing confirmed cases and the highest confirmed cases in China and Hubei Province have

| Table9: Confirmed cases (cumulative)/表9:确诊病例（累计值） | National/全国 | Daily Change in National/全国日度变化 | Hubei/湖北 | Daily Change in Hubei/湖北日度变化 | Wuhan City/武汉 | Daily in Wuhan/武汉日度变化 | Xiaogan City/孝感 | Daily Change in Xiaogan/日度变化 | Suizhou City/随州 | Dialy Change in Suizhou/随州日度变化 | Huanggang city/黄冈 | Daily Change in Huanggang/日度变化 | Huangshi City/黄石 | Daily Change in Huangshi/黄石日度变化 |
|---|---|---|---|---|---|---|---|---|---|---|---|---|---|---|
| 2020/2/12 | 58,761 | 31.6% | 47,163 | 41.4% | 32,081 | 64.0% | 2,874 | 4.5% | 1,160 | 2.7% | 2,628 | 9.6% | 911 | 4.2% |
| 2020/2/13 | 63,851 | 8.7% | 51,986 | 10.2% | 35,991 | 12.2% | 3,009 | 4.7% | 1,206 | 4.0% | 2,791 | 6.2% | 943 | 3.5% |
| 2020/2/14 | 66,492 | 4.1% | 54,406 | 4.7% | 37,914 | 5.3% | 3,114 | 3.5% | 1,232 | 2.2% | 2,817 | 0.9% | 980 | 3.9% |
| 2020/2/15 | 68,500 | 3.0% | 56,249 | 3.4% | 39,462 | 4.1% | 3,201 | 2.8% | 1,254 | 1.8% | 2,823 | 0.2% | 988 | 0.8% |
| 2020/2/16 | 70,548 | 3.0% | 58,182 | 3.4% | 41,152 | 4.3% | 3,279 | 2.4% | 1,267 | 1.0% | 2,831 | 0.3% | 983 | -0.5% |
| **2020/2/17** | **72,436** | **2.7%** | **59,989** | **3.1%** | **42,752** | **3.9%** | **3,320** | **1.3%** | **1,278** | **0.9%** | **2,828** | **-0.1%** | **985** | **0.2%** |
| **2020/2/18** | **74,182** | **2.4%** | **61,682** | **2.8%** | **44,412** | **3.9%** | **3,344** | **0.7%** | **1,280** | **0.2%** | **2,844** | **0.6%** | **983** | **-0.2%** |
| **2020/2/19** | **74,576** | **0.5%** | **62,031** | **0.6%** | **45,027** | **1.4%** | **3,329** | **-0.4%** | **1,283** | **0.2%** | **2,839** | **-0.2%** | **967** | **-1.6%** |
| **2020/2/20** | **75,891** | **1.8%** | **63,088** | **1.7%** | **45,346** | **0.7%** | **3,427** | **2.9%** | **1,292** | **0.7%** | **2,883** | **1.5%** | **992** | **2.6%** |
| **2020/2/21** | **76,288** | **0.5%** | **63,454** | **0.6%** | **45,660** | **0.7%** | **3,429** | **0.1%** | **1,296** | **0.3%** | **2,899** | **0.6%** | **997** | **0.5%** |
| **2020/2/22** | **76,936** | **0.8%** | **64,084** | **1.0%** | **46,201** | **1.2%** | **3,443** | **0.4%** | **1,300** | **0.3%** | **2,904** | **0.2%** | **1,001** | **0.4%** |

Table10: The Existing Confirmed Patients/表10：全国现存确诊病例变化时间表

| Time/时间 | National # of Existing Confirmed Patients/ 全国确诊病例存量 | Daily Change in National/全国日度变化 | # of Existing Confirmed Patients in Hubei/湖北确诊病例存量 | Hubei Daily Change/湖北日度变化 |
|---|---|---|---|---|
| 2020/2/11 | 38,800 | | 29,659 | |
| 2020/2/12 | 51,860 | 33.7% | 42,789 | 44.3% |
| 2020/2/13 | 55,748 | 7.5% | 46,806 | 9.4% |
| 2020/2/14 | 56,872 | 2.0% | 48,175 | 2.9% |
| 2020/2/15 | 57,416 | 1.0% | 49,030 | 1.8% |
| 2020/2/16 | 57,927 | 0.9% | 49,847 | 1.7% |
| **2020/2/17** | **58,016** | **0.2%** | 50,338 | **1.0%** |
| **2020/2/18** | 57,802 | **-0.4%** | **50,633** | **0.6%** |
| **2020/2/19** | 56,727 | -1.9% | 50,091 | -1.1% |
| **2020/2/20** | 55,389 | -2.4% | 49,156 | -1.9% |
| **2020/2/21** | 53,285 | -3.8% | 47,647 | -3.1% |
| **2020/2/21** | 51,606 | -6.8% | 46,439 | -5.5% |

been reduced by more than 5,700 and 5,000 respectively. Considering the three-day screening in Hubei Province from Feb.17, to 19, 2020, we have reason to believe that the true Turning (Inflection) Point based on the number existing confirmed cases emerged at approximately Feb. 17, 2020.

2) Through a comprehensive analysis of Hubei province, we determine the traditional inflection point to be between Feb. 17, and Feb. 19, 2020. This data set combined with the previous conclusion lead us to believe that the COVID-19 infection in China (including Wuhan, Hubei) has been fully controlled since mid-February 2020, with Feb. 18 as the most precise date (we arrive at Feb. 18 as after Feb. 17, the daily change of confirmed patients is less than 1% in negative numbers).

## 6. Final Concluding Remarks based on our COVID-19 Predictions in China

We believe that our iSEIR model is a reliable predictor for the Turning Period of the struggle against COVID-19 through introducing concepts of "Turning Phase" and "Supersaturation Phenomenon". This model accounts for intervention policies and methods such as isolation control programs (e.g. quarantine); such as those implemented in February 2020 in China (see [33] - [34]). Beyond using our iSEIR model for the COVID-19 in China from late December to early March 2020, we hope to further apply it to outbreaks worldwide.

When we incorporate the public official data released by the Chinese government since January 23, 2020, our iSEIR simulation allowed us to conclude that:

- February 1, 2020, is the starting point for the formation of the turning point of the COVID-19 epidemic situation in China. Considering the intervention efforts implemented at the state-level, the time period for the control of the COVID-19 epidemic should be around mid-February 2020, and at least by the end of February 2020 (see the original report released [30]-[32] for more details).

Before arriving at our final remarks, we want to consider the following with regards to the current global state of the COVID-19 outbreak:

1) The virus can be spread by infected and asymptomatic individuals;
2) At current, there is no fully effective medicine or treatment; and
3) No vaccine.

With these conditions, it is important that our iSEIR model be used to aid in modeling the timeline of the COVID-19 outbreak in other countries in shaping public policy. However, we acknowledge that our model is patterned to reflect effective intervention methods and protocols such as the "Wuhan Quarantine" (see [36]). In conclusion, the simulated timeline provided by our iSEIR model best fits outbreak scenarios where early adoption of public health controls and restrictions are implemented to flatten the curve.

Finally, we want to reinforce that emergency risk management is always associated with the implementation of an emergency plan. The identification of the Turning Time Period is key to emergency planning as it provides a timeline for effective actions and solutions to combat a pandemic by reducing as much unexpected risk as soon as possible. We can further improve our ability to emergency-plan in urgent events, such as in the control of an infectious disease outbreak like COVID-19 in three key areas:

1) a better spatiotemporal model to describe the mechanics of the spread of infectious diseases;
2) an efficient way to conduct numerical simulation to identify the "Turning Time Period (Phase)" for the emergency event, e.g., the timeframe for the outbreak of infectious diseases spread such as OVID-19 virus; and

3) carrying out effective predictive analysis by establishing a coherent bigdata method for data fusion from different sources with different structures to support a dynamic management to respond to daily issues with effective emergency planning.

## Appendix I: The General Inputs for the Simulation of iSEIR Model

| **The General Inputs for the Simulation under the framework of iSEIR Model** | |
|---|---|
| Date: February 10, 2020.<br>The data in highlight with yellow color means on the true real scene for the simulation | |
| Name of Parameter/参数名称 | Values/取值 |
| *# of simulations* | 100 times of simulations |
| *N: # of groups/communities* | Specified in simulation, |
| *M: # of cities* | 10 |
| *ρ: distribution density* | 0.4 |
| *c: distance* | base on Euclidean distance formula/用欧式距离公式计算 |
| *T: time steps* | *1~3000 (one unit is around 10 minutes, or 5 minutes)* |
| *N: population* | The input based on the true real scene for the simulation |
| *i, j, k : random individual* | Uniform distribution |
| *i, j, k:* | The range is from N specified by true real scene |
| $\gamma$ | The input based on the true real scene for the simulation |
| *A* | = 0.000001 |
| $p_{ij}$ | = 1, if the distance of i and j < c;<br>= 0; otherwise. |
| $q_{jk}$ | = 1, if the distance of j and k < c;<br>= 0, otherwise. |
| $\mu_i$ | = 1, if in [0.0001, 1];<br>= 0, otherwise. |
| $\varepsilon_i$ | = 1, if <=0.0001;<br>= 0, otherwise. |
| $\beta_j$ | = 1, if in [0.001, 1];<br>= 0, otherwise. |
| $\alpha_j$ | = 1, if <=0.001;<br>= 0, otherwise |
| $\lambda_k$ | = 1, if <= the specified value for a given input parameters ($\gamma$)<br>= 0, otherwise. |
| *Other parameters* | Based on the situation to specify |